\documentclass[fleqn,usenatbib]{mnras}

\usepackage{newtxtext}

\usepackage[T1]{fontenc}
\usepackage{ae,aecompl}
\usepackage{multirow}
\usepackage{array}

\usepackage[normalem]{ulem}

\usepackage{graphicx}	
\usepackage{amsmath}	
\usepackage{amssymb}	

\newcommand{\mlike}{\mathcal{M}}

\newcommand{\drxn}{\boldsymbol{\omega}}
\newcommand{\drxnp}{\boldsymbol{\omega}'}
\newcommand{\drxnpp}{\boldsymbol{\omega}''}
\newcommand{\x}{\boldsymbol{x}}

\newcommand{\C}[1]{\texttt{C#1}}

\newcommand{\rota}[1]{R_{#1}}
\newcommand{\rot}{\rota{\phi}}

\newcommand{\be}{\begin{equation}}
\newcommand{\ee}{\end{equation}}

\title{Optimal Probabilistic Catalogue Matching for Radio Sources}

\author[Dongwei Fan et al.]{
Dongwei Fan$^{1,4,7}$\thanks{E-mail:fandongwei@nao.cas.cn},
Tam\'as Budav\'ari$^{2,3,4}$, 
Ray P. Norris$^{5,6},$ 
Amitabh Basu$^{2,3}$
\\
$^{1}$National Astronomical Observatories, Chinese Academy of Sciences, 20A Datun Road, Beijing 100101, P.R.China\\
$^{2}$Department of Applied Mathematics and Statistics, The Johns Hopkins University, 3400 North Charles Street, Baltimore, MD 21218, USA\\
$^{3}$Department of Computer Science, The Johns Hopkins University, 3400 North Charles Street, Baltimore, MD 21218, USA\\
$^{4}$Department of Physics and Astronomy, The Johns Hopkins University, 3400 North Charles Street, Baltimore, MD 21218, USA\\
$^{5}$Western Sydney University, Locked Bag 1797, Penrith, NSW 2751, Australia\\
$^{6}$CSIRO Astronomy \& Space Science, P.O. Box 76, Epping, NSW 1710, Australia\\
$^{7}$National Astronomical Data Center, Beijing 100101, P.R.China
}%
\date{Accepted XXX. Received YYY; in original form ZZZ}

\pubyear{2020}

\begin{document}
\label{firstpage}
\pagerange{\pageref{firstpage}--\pageref{lastpage}}
\maketitle

\begin{abstract}
Cross-matching catalogues from radio surveys to catalogues of sources at other wavelengths is extremely hard, because  radio sources are often extended, often consist of several spatially separated components, and  often no radio component is coincident with the optical/infrared host galaxy. Traditionally, the cross-matching is done by eye, but this does not scale to the millions of radio sources  expected from the next generation of radio surveys. We present an innovative  automated procedure, using Bayesian hypothesis testing, that models trial   radio-source morphologies with putative positions of the host galaxy. This new algorithm differs from an earlier version by allowing more complex radio source morphologies, and performing a simultaneous fit over a large field. We show that this technique performs well in an unsupervised mode.
\end{abstract}




\section{Introduction}\label{introduction}

Techniques for automated cross-matching catalogues of optical and infrared (IR) sources are well-established, resulting in services such as Vizier (Ochsenbein,
Bauer \& Marcout 2000),  Simbad (Wenger et al. 2000),  NASA
Extragalactic Database (NED; Mazzarella, Madore \& Helou 2001),
NASA/IPAC Infrared Science Archive (IRSA), and SkyQuery (Malik
et al. 2003; Budavari, Dobos \& Szalay 2013).

Cross-matching radio sources to optical/IR catalogues is much more difficult, because about 10\% of radio sources consist of extended radio lobes resulting from the interaction of relativistic jets of electrons emitted from the environs of a super-massive black hole (SMBH) in the nucleus of the host galaxy. Sometimes this central core is visible in the radio, resulting in a triple radio source, and sometimes only the lobes are visible, resulting in a double radio source.

For clarity in this paper, we distinguish between a radio ``component'', consisting of a single blob of radio emission, and a radio ``source'', consisting of all the radio emission associated with one host galaxy. In general, one radio source will consist of several radio components.

One consequence is that a pair of radio components might either be two separate galaxies, or a pair of radio lobes surrounding one galaxy. Often, this question can only be resolved by comparing the radio image with the optical/IR image, so that the process of cross-identification merges with the process of classification.

Another consequence is that there is often no radio component coincident with the host galaxy, so that associating radio catalogues with optical/IR catalogues needs to take account of the overall radio source morphology, ruling out the use of simple nearest-neighbour of likelihood-ratio algorithms \citep[e.g.,][]{weston18}.
As a result, this cross-matching has traditionally been done by eye, but this becomes impractical for future radio surveys in which tens of millions of objects are likely to be discovered \citep{norris17}. This present paper is primarily driven by the Evolutionary Map of the Universe (EMU) project \citep{norris11} which hopes to catalogue about 70 million radio sources.

One response to this challenge has been the development of Radio Galaxy Zoo \citep{banfield15}, in which citizen scientists cross-match radio sources with their infrared counterparts. Another response has been the development of the rapidly growing field of machine-learning (ML) approaches to classifying and cross-matching radio and IR/optical sources \citep[e.g.,][]{aniyan17, alger18, lukic18, wu19, galvin19, ralph19}.

As an alternative to ML approaches, we presented \citep{fan15}  a Bayesian approach in which, for each source, we constructed a number of hypotheses. In each hypothesis we identified individual radio components as lobes or core sources, and trialled putative associations with the host galaxy, chosen from a catalogue of infrared sources. We then chose the hypothesis for each source that had the highest probability.
The ILP differs from ML techniques in that it uses prior expert knowledge about the morphology of radio galaxies, whereas supervised ML techniques \citep[e.g.,][]{wu19} use the knowledge derived from training sets, and unsupervised ML techniques \citep[e.g.,][]{galvin20} use no prior knowledge but rely on large samples to establish the common morphologies. It is currently unclear which technique will be most effective for large radio survey. Instead, all three approaches (ILP, supervised ML, unsupervised ML) need to be tested on real large data sets.

In this paper, we extend the algorithm of \citet{fan15} in two ways:
(1) We use a more sophisticated radio source model in which we allow the line connecting the lobes of the source with the nucleus to be bent.
(2) Having constructed a number of hypotheses for each source, we then choose a solution that maximizes the likelihood over the whole field, rather than focusing on the likelihood for each source separately.
In this paper we call the earlier algorithm used by \citet{fan15} the ``greedy'' approach, and we call the present approach ``ILP'', for Integer Linear Programming.

\section{Matching for Radio Sources}
\label{sec:method}

In \citet{fan15} we introduced the Bayesian formalism for including  realistic morphology of a radio source into the cross-identification process. We evaluate the \textit{marginal likelihood} of the competing hypotheses, namely with possible configurations describing a given set of radio sources and their counterparts. Conceptually, one considers any combination of radio and other objects to (numerically) compute the probability of the data given those assumptions.
For example, if we are looking at a possible association of two radio detections and one infrared source, it could be that we see the core in both the radio and infrared,  and
a single lobe in the radio, or it could be that the infrared source is the core but the radio shows two lobes.
But the list of possible hypotheses is much longer: maybe all sources are from a separate source or the lobes are from one object but the infrared is another object and so on. Our computer program iterates through every possible combination and evaluates an objective quality, the marginal likelihood, for each.

Here we make two improvements to our previously successful automated procedure. First, we explicitly allow for morphology where the lobes are not along a straight line connected through the core.
This requires us to introduce a prior on the possible angles but the model is more realistic than
simply assuming they fall along a straight line as before.
The beauty of the Bayesian treatment is that the marginal likelihoods of the competing hypotheses are well-defined for this more advanced model, and so we simply have to marginalize over one more parameters.
The second and most crucial improvement comes from a global optimization that considers the matched catalogue as a whole instead of using a greedy procedure to pick out the associations.

\subsection{Flexible Model for Radio Morphology}

For any given hypothesis, we introduce the true position of the assumed latent object (i.e., the host galaxy).
We consider every possible position of the host galaxy relative to the radio components, and let the available data (measured positions and their uncertainty) provide appropriate weighting via the likelihood function, e.g., a Gaussian.
The computational methods for efficient evaluations are discussed in \citet{fan15} in detail.

\begin{figure}
\begin{center}
\includegraphics[width=0.5\textwidth]{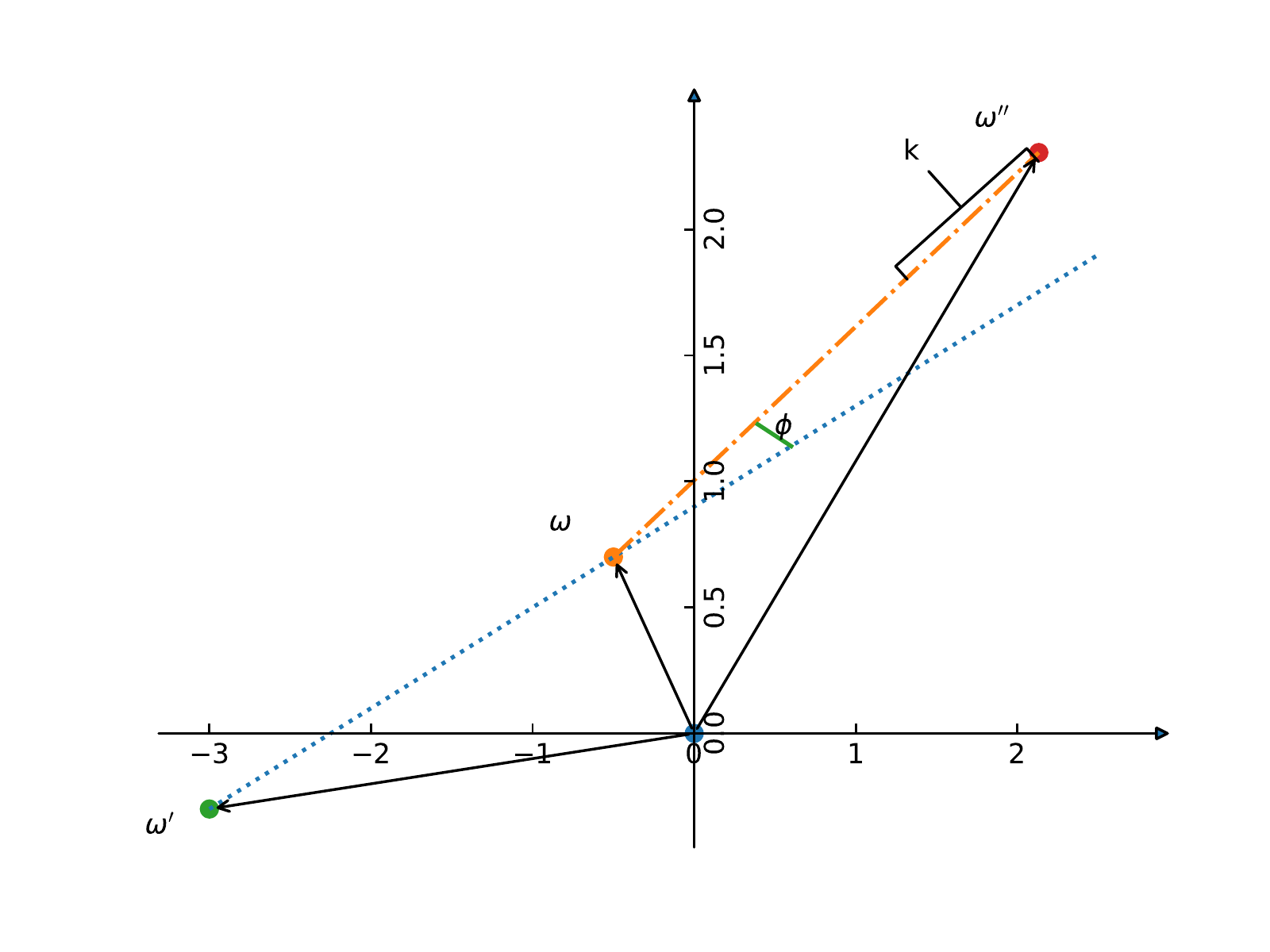}
\end{center}
\caption{Geometry model for the radio morphology includes the position or all assumed sources at varying distances and angles, see text.
Blue point at the coordinates origin represents the infrared source, while yellow point above the origin is radio core $\drxn$, green point at left bottom represents radio lobe 1 $\drxnp$, red dot at the top right is radio lobe 2 $\drxnpp$. $\phi$ is the angle between lines $\drxn-\drxnp$ and $\drxnpp-\drxn$.
\label{fig:rmodel}}
\end{figure}

The updated model is illustrated in Figure~\ref{fig:rmodel} in the tangent plane.
For the calculations, we chose the position of the infrared detection as the origin. In this 2D coordinate system, the radio core is at $\drxn$ and the lobes are at $\drxnp$ and $\drxnpp$. The following equations describe their relations to each other using additional model parameters $\phi,k$: $k$ allows for the lobes to be at different distances from the core, and $\phi$ permits (slight) bending.
\begin{equation}
\drxnpp = \drxn + \rot\,\big[(1\!+\!k)\,(\drxn-\drxnp)\big]
\label{eqn:angle2}
\end{equation}
where $\rot$ is the usual 2$\times$2 matrix of the rotation by $\phi$.
All model parameters $\theta = \left\{\drxn, \drxnp, k, \phi\right\}$ are involved in the assessment of the associations along with their priors, which we can write as
\begin{equation}\label{eq:prior}
\rho(\theta) = \rho(\drxn)\,\rho(\drxnp|\drxn)\,\rho(k)\,\rho(\phi)
\end{equation}
due to the appropriate (in)dependencies. Hereafter we assume that $\rho(\drxn)$ is isotropic (or uniform on the surface of the sphere) and $\rho(\drxnp|\drxn)$ depends only on the separation of the two positions. Furthermore, we will assume simple Gaussian priors for $k$ and $\phi$ with 0 mean, a combination that corresponds to a straight and symmetric configuration of the two lobes.

\subsection{Matched Catalogues as Partitions of Detections}
\label{partitions}

Following the approach of \citet{budavari15} and \citet{shi2019} we ideally consider all possible matched catalogues with every possible combinations of associated detections.
Let $(i,c)$ represent the $i$th detection (or an internal identifier) in catalogue $c$ with measured position $\x_{ic}$. The collection of all detections $D$ is the set of all $(i,c)$ measurements. We adopt the use of the \citet{fisher} distribution for the likelihood calculation,
\begin{equation} \label{eq:fisher}
f\left(\x;\drxn,\kappa\right)=\dfrac{\kappa}{4 \pi \sinh \kappa} \ \exp\big(\kappa\, \drxn \x\big)
\end{equation}
that essentially corresponds to a Gaussian in the usual flat sky limit for small uncertainty, where the compactness parameter $\kappa=1/\sigma^2$, and the $\sigma$ is the positional error from the catalogue.

The final cross-matched catalogue will consist of multiple associations, which we will call objects, and index them by $o$.
A particular object $o$ will have an $S_o$ set of $(i,c)$ detections in its association.
In fact, we can think of matching radio observations as if the radio catalogue appeared multiple times according to what roles the radio detections play in a particular association: a core or one of the two possible lobes.
So, for matching IR and radio observations, we order the collections as follows: \mbox{$c\!=\!0$} corresponds to the IR catalogue,
\mbox{$c\!=\!1$} represents radio core detections,
and
\mbox{$c\!=\!2, 3$} correspond to lobes.
Conceptually, the catalogues corresponding to \mbox{$c\!=\!1$}, $2$, and $3$ are identical copies of the radio observations. In practice, however, they really are just views of the catalogue; there is no need for duplication of the data.
Naturally, a given radio detection can only appear in one association $o$, which we will enforce explicitly when looking for the optimal catalogue.

If the measurements in $S_o$ correspond to one object, the marginal likelihood of that $o$ is calculated by the integral
\begin{equation}\label{eq:Mo}
\mlike_{o}=\int\!\!d\theta \ \rho_{o}\!\left(\theta\right)\!\prod_{\left(i,c\right)\in S_{o}}\!\!\ell_{ic}(\theta) \\
\end{equation}
where the prior $\rho_o$ has an index to signify the possibility that different catalogues could have different sky coverage, hence different overlapping regions would yield different priors, see discussion in \citet{pxid} and \citet{budavari15}.
The member likelihoods are simply given by the Fisher distribution at the appropriate true positions,%
\begin{equation}
\ell_{ic}(\theta) = \left\{
    \begin{array}{ll}
    f(\x_{ic};\,\drxn\,,\,\kappa_{ic}) & \textrm{if}~c=0~\textrm{or}~1 \\
    f(\x_{ic};\,\drxnp,\,\kappa_{ic}) & \textrm{if}~c=2\\
    f(\x_{ic};\,\drxnpp, \kappa_{ic}) & \textrm{if}~c=3
    \end{array}
\right.
\end{equation}
where $\drxnpp$ is really a function of all parameters, as given by eq.(\ref{eqn:angle2}) above.
To find the optimal catalogue, we need to find the set of objects such that the product of their marginal likelihoods, $\prod \mlike_o$, is maximal.

In addition, we also introduce the marginal likelihood of ``no association'', where each detection is considered by itself as a separate object. The integrals simplify to the familiar product
\begin{equation}
\mlike_{o}^{\textrm{NA}}= \prod_{\left(i,c\right)\in S_{o}} \int\!\!d\omega\ \rho_{c}\!\left(\omega\right)\,\ell_{ic}\!\left(\omega\right)
\end{equation}
with which we can define a Bayes factor for object $o$ as
\begin{equation}\label{BayF}
B_{o}\equiv \dfrac{\mlike_{o}}{\mlike_{o}^{\textrm{NA}}}\,.
\end{equation}
Considering that the product of the marginal likelihoods $\prod \mlike_{o}^{\textrm{NA}}$ over all $o$ in a matched catalogue always contains the same terms corresponding to every $(i,c)$ detection, the product is constant across every possible partitioning. This means that the optimal catalogue will also have maximal $\prod B_o$ value.
Using the latter over the product of marginals has the advantage that \mbox{$B_o\!=\!1$} for the \textit{orphan} objects that consist of a single detection, and hence their contribution to the product can neglected.

\subsection{Combinatorial Optimization}
\label{sec:comb-opt}

To formalize the above maximization of the global cross-match catalogue likelihood \citep{budavari_basu}, one first has to enumerate all possible associations. Following \citet{shi2019} we index these by $T$ and introduce a binary variable $x_T$ that is 1 if the association is selected for the cross-match catalogue and 0 otherwise.
Since a radio detection should be used only from exactly one of the three catalogues $c=1,2,3$, we must not have any variables that correspond to a set of detections where the same detection is used from more than one catalogues $c\in \{1,2,3\}$. More precisely, we do not include a variable for any set $S_o$ which contains $(i,c)$ {\em and} $(i',c')$ such that \mbox{$i\!=\!i'$} but \mbox{$c\!\neq\!c'$}.

The Bayes factors can be (numerically) evaluated for every one of the candidates, $B_T$. Formally, the optimization over \mbox{$x\!=\!\{x_T\}$} can now be written as
\be
\min_x \sum_T w_T x_T
\ee
where \mbox{$w_T\!=\!-\log B_T$}, but further constraints are required to ensure that every detection appears exactly once in the resulting catalogue, and each radio detection appears only in one of the three possible roles \mbox{($c\!=\!1$, $2$, $3$)}. Considering that orphans have no contributions to the above objective as their \mbox{$w_T\!=\!0$}, we can write inequalities that are better handled by optimization algorithms. For example,
\be
\sum_{T\ni{}\,(i,c)} \!\!\! x_T \leq 1 \qquad \forall\,(i,c) \in D
\ee
means that every $(i,c)$ detection can appear at most in one $T$ association of a matched catalogue: the sum goes over all $T$ that include $(i,c)$.
For the radio detections, we have further constraints to ensure that each and every one of them only appears (at most) in one role, i.e., $c=1$, $2$ or $3$, which we can write as
\be
  \sum_{T\ni{}\,(i,1)} \!\!\! x_T
+ \sum_{T\ni{}\,(i,2)} \!\!\! x_T
+ \sum_{T\ni{}\,(i,3)} \!\!\! x_T
\leq 1 \qquad \forall\, i \in R 
\ee
where $R$ is the set of indices of the detections in the radio catalogue, i.e., \mbox{$R\!=\!\left\{\,i: (i,\!1)\!\in\!D\,\right\}$}.
Such an optimization problem falls in the realm of Integer Linear Programming (hereafter ILP) for which various commercial and research solvers are available off the shelf.

\subsection{Priors on Matched Catalogues}

The above optimization considers only the marginal likelihoods of the associations, which neglects any prior information about the resulting catalogues.
If we had such knowledge, the objective could be easily updated to include that.
For example, one might have a distribution of the expected number of objects in a catalogue, or perhaps know the relative frequency of occurrence for the different kinds of matches, such as radio triples, doubles, etc.
We expect that with more data, a hierarchical modeling approach will be eventually developed to simultaneously learn the population distributions and the assignments, but for now we will resort to the simplest possible scenario:
In order to maximise the number of radio components in each association, we minimize the number of final associations. In case that, a group of three radio components is more likely to be associated with a single galaxy, than three separate galaxies, if these two scenarios each have the same calculated probability.

Our modified optimization would now have two terms,
\be\label{eq:multi-obj}
\min_x \left[ \sum_T w_T x_T + \lambda \sum_T x_T \right]
\ee
where $\lambda$ expresses the strength of our preference for fewer objects. This objective is also linear, and hence does not add to the complexity of the problem.
Our input data  consists only of the directions of the detections, with inherent uncertainties, and ignores other information that could in principle be used as constraints.
For example, we expect that additional shape and flux measurements will play a key role in defining scientific catalogues, but these need to be studied further with more data.
With these caveats and ideas in mind, we will explore solutions using the multi-objective approach, as in eq.\eqref{eq:multi-obj}.

\section{Application to SWIRE \& ATLAS CDFS}

Here we apply our new approach to two catalogues discussed by \citet{norris06}: (1) the Spitzer Wide-area InfraRed Extragalactic Survey~\citep[SWIRE;][]{lonsdale} over Chandra Deep Field South (CDFS): SWIRE CDFS Region Fall '05 Spitzer Catalogue\footnote{SWIRE CDFS Region Fall '05 Spitzer can be searched and downloaded on \url{https://irsa.ipac.caltech.edu/cgi-bin/Gator/nph-scan?mission=irsa&submit=Select&projshort=SPITZER}}
and (2) the Australia Telescope Large Area Survey (ATLAS) of the CDFS.
The SWIRE catalogue contains 221,535 entries, but for consistency and a fair comparison, we append 119 additional detections per the work of \citet{norris06}; see their Table~6 and the discussion therein.
The ATLAS collection lists 784 radio components, which appear in Table~4 in~\citet{vizier51322409}.

\subsection{Practical Considerations}
\label{practical}

The cross-matching process consists of several steps.
The first step is to remove highly unlikely associations,  using a spatial two-way matching between the IR and radio catalogues.
This helps in the efficiency of the method by reducing the number of variables $x_T$ in the optimization (see the discussion in Section~\ref{sec:comb-opt}).
In particular, when two measurements are far from each other on the sky, they are unlikely to be related.
We set the angular separation threshold for the preprocessing to 2\arcmin, which is a safe limit considering that the farthest components in the radio triples associated by \citet{norris06} are 61.5\arcsec apart.

In the second step, all possible hypotheses are enumerated as possible  candidates for associations.
When considering two lobes we break the degeneracy by considering the farthest lobe to be at $\drxnp$ and the closer one at $\drxnpp$, then we evaluate the marginal likelihoods for all combinations of detections.
For the member likelihood functions, we choose $\sigma_0\!=\!0.2$\arcsec\ astrometric uncertainty for SWIRE, $\sigma_1\!=\!1.2$\arcsec\ for the radio cores and
$\sigma_2\!=\!\sigma_3\!=\!2.2$\arcsec\ for the lobes, whose position is harder to determine due to their more extended and somewhat asymmetric nature.

The prior probability density function in eq.\eqref{eq:prior} is assumed to be isotropic over $\drxn$ (see discussions in \citet{pxid} and \citet{budavari15}).
The conditional PDF $\rho(\drxnp \vert \drxn)$ is assumed to be a function of only the angular separation of $\drxn$ and $\drxnp$, and its radial dependence has the shape of the Rayleigh distribution with $\sigma_R=8.5"$.
The dimensionless parameter $k$ that describes the difference in the lobe separations from the core is assumed to have a Gaussian prior with a standard deviation of \mbox{$\sigma_k\!=\!1$}.
The angle $\phi$ that describes how bent the geometry is also assumed to be a Gaussian with 0 mean and \mbox{$\sigma_{\phi}\!=\!15^{\circ}$} degrees.
The 15 degrees is training from \citet{norris06}, of which maximal angle is 16.96 degrees.

\begin{figure}
\begin{center}
\includegraphics[scale=0.55]{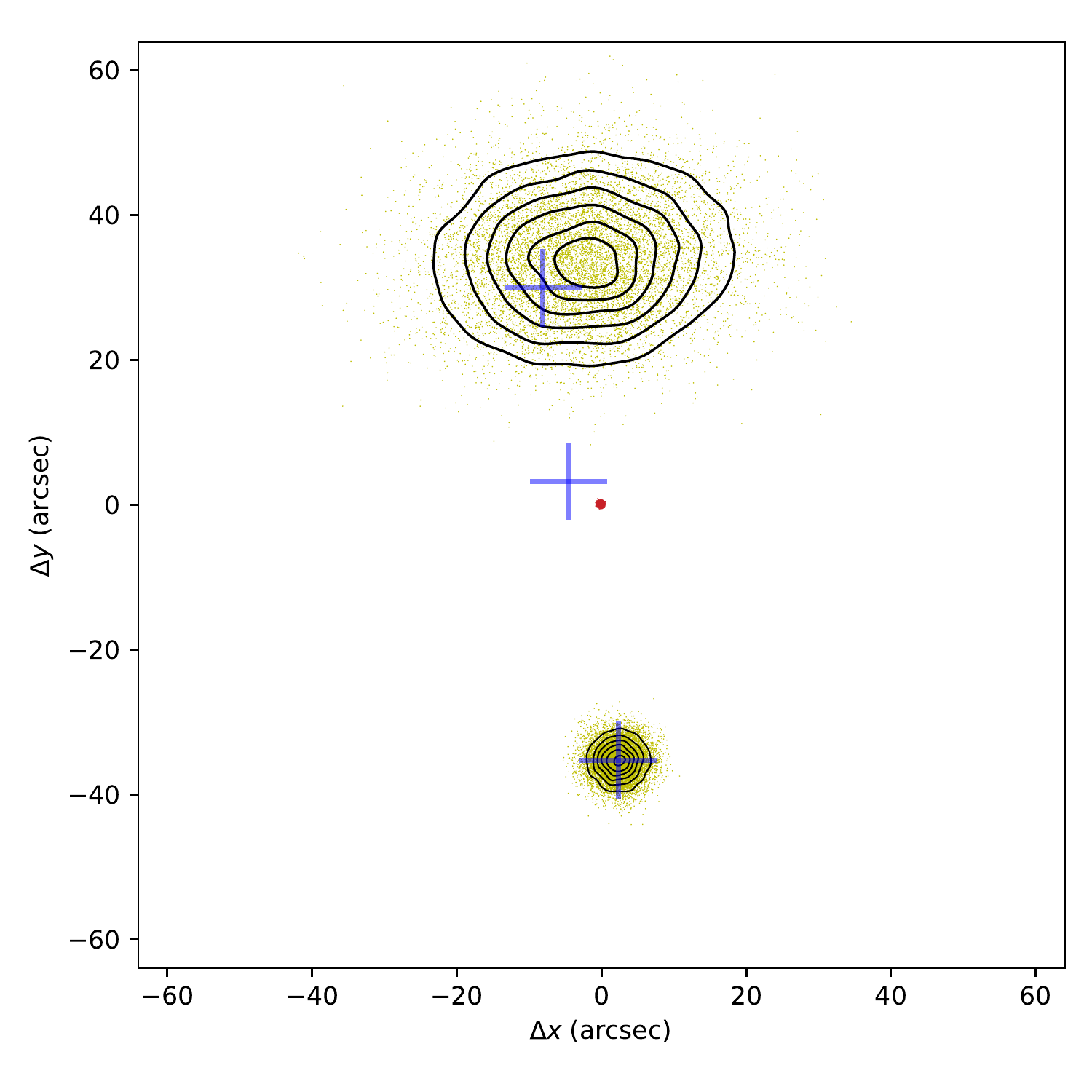}
\end{center}
\caption[Caption]%
{Samples in the numerical marginalization over 10,000 points of a radio triple. For efficiency, we sample from the likelihoods using importance sampling, as described in the text.
The three blue crosses represent the positions of the two radio lobes and the radio core. The red spot at the center shows the position of the IR source, corresponding to $\omega$. The yellow points and black contours sample the positions of the two radio lobes, corresponding to $\omega^\prime$ and $\omega^{\prime\prime}$. Contours illustrate the sample points' density using Gaussian kernels density estimation\protect\footnote{gaussian\_kde,  \url{https://docs.scipy.org/doc/scipy/reference/generated/scipy.stats.gaussian\_kde.html}, as described in the text.}
}
\label{img:sample-triplets}
\end{figure}

With those in hand, our implementation numerically integrates the marginal likelihoods and computes the Bayes factors.
For an efficient integration, we sample $\drxn$ from the IR core's likelihood function and $\drxnp$ from the farthest radio lobe's likelihood. Given those and random samples from the Gaussian priors for $k$ and $\phi$, we can generate a random $\drxnpp$ and evaluate the contribution of that sample to the integral.
Figure \ref{img:sample-triplets} illustrates 10,000 sample points for a known radio triple, whose positions are shown by large blue crosses. Red contours around (0,0) represent the density of $\drxn$ samples, which are highly concentrated due to the small uncertainty of the IR source, and therefore appear as a red dot at the origin.
The core radio component (the blue cross near the origin) is located a small distance from the origin due to its larger uncertainty -- its likelihood is evaluated at $\drxn$. The black contours and the yellow points toward the bottom of the figure show the $\drxnp$ samples from the likelihood function of the radio lobe.
The contours on top for $\drxnpp$ show the samples that are a combination of all model parameters, $\drxn$, $\drxnp$, $k$ and $\phi$.

When every \mbox{$w_T\!=\!-\log B_T$} is computed, we solve the minimization using the ILP implementation by \citet{gurobi} for the binary variables $x_T$ using the aforementioned objectives and constraints.
Gurobi has builtin routines to handle the multi-objective optimization of eq.\eqref{eq:multi-obj}, for which we adopt \mbox{$\lambda\!=\!2.5$} to prefer solutions with fewer objects, i.e., larger associations of detections.
Considering that the average weight $\bar{w}$ is significant for selected associations, the weighting effectively is by a \mbox{$\lambda'=\lambda/\lvert\bar{w}\rvert$} which in our experiments is about one fifth.

\subsection{Results and Discussion}
\label{results}
First, some caveats: \citet{norris06} associate the SWIRE and ATLAS measurements by eye. Expert radio astronomers look at the IR images on top of which multiple contours show the details of radio intensity, and they annotate the catalogues based on such rich views of the data.
In comparison, our current approach considers only the positions of the detections and their uncertainty but not their brightness nor their shapes.
We consider the expert matched catalogue as our ground truth, hereafter called the ``reference  catalogue'', to which we compare our results. We hope that our approach can approximately recover many of the known associations without introducing too many spurious matches.
However, we also caution that in some cases the classifications are ambiguous, and even experts will sometimes disagree over the correct classification, so the ``reference  catalogue'' is not always reliable.
The reference catalogue primarily consists of IR detections associated with radio triples, doubles and singles but also lists six complex associations, see Table~\ref{tbl:6complex}, which do not fit the usual model. We will discuss all these scenarios in turn.

\begin{table}
\begin{center}
\caption{Results compared to the reference catalogue.
Column 2 shows the number of each class found by our algorithm.
Column 3 shows the number of each class found both in reference  catalogue and by our algorithm.
Column 4 shows the number of each class that were in reference  catalogue but missed by our algorithm.
Column 5 shows the number of each class that were found by our algorithm but were not in reference  catalogue.\label{tbl:ilp-sols} }
\begin{tabular}{lrrrr}
\hline\hline
Association type & Found & Common & Missed & Extra\\
\hline
Radio triples & 12 & 10 & 0 & 2 \\
Radio doubles & 31 & 21 & 4 & 10 \\
Radio singles  & 684 & 680 & 5 & 4 \\
\hline\hline
\end{tabular}
\end{center}
\end{table}

\begin{table*}
\begin{center}
\caption{Six complex combinations in \citet{norris06}\label{tbl:6complex} visual matching}
\begin{tabular}{lccll}
\hline\hline
Cplx & SID & SWIRE designated name & Radio Components & Comment from the Visual Inspection\\
\hline
\#1 & \texttt{S136} & \texttt{SWIRE3\_J032825.92-271701.3} & \C{141}, \C{148}, \C{151} & \textit{Radio double with connecting jet}\\
\#2 & \texttt{S349} & \texttt{(336555)} & \C{366}, \C{369}, \C{376} & \textit{complex jet structure}\\
\#3 & \texttt{S409} & \texttt{SWIRE3\_J033210.74-272635.5} & \C{437}, \C{440}, \C{446} & \textit{weird complex source with tails. z(g)}\\
\#4 & \texttt{S631} & \texttt{SWIRE3\_J033437.35-272652.2} & \C{679}, \C{681}, \C{683} & \textit{Radio double}\\
\#5 & \texttt{S707} & \texttt{SWIRE3\_J033533.90-273310.9} & \C{759}, \C{760}, \C{761}, \C{764} & \textit{Centre of a linear complex (jets or grav arc?)}\\
\#6 & \texttt{S719} & \texttt{SWIRE3\_J033542.52-274344.1} & \C{774}, \C{775}, \C{777} & \textit{Radio double}\\
\hline\hline
\end{tabular}
\end{center}
\end{table*}

Our implementation produces an optimal cross-matched catalogue under the morphology model described in Section~\ref{sec:method}.
Table~\ref{tbl:ilp-sols} lists the best solution along with the number of discovered associations with radio triples, doubles and singles.
In fact, we separate them into three categories based on whether they appear in the reference catalogue as common, missed, or extra.
We see that our code finds all 10 of the radio triples in the reference catalogue with only two extra associations.
Out of the 25 radio doubles we find 21 with an additional 10 extra.
To determine how good these results really are, we now examine in turn the associations with radio triples, doubles and singles.

\begin{figure}
\begin{center}
\includegraphics[scale=0.2]{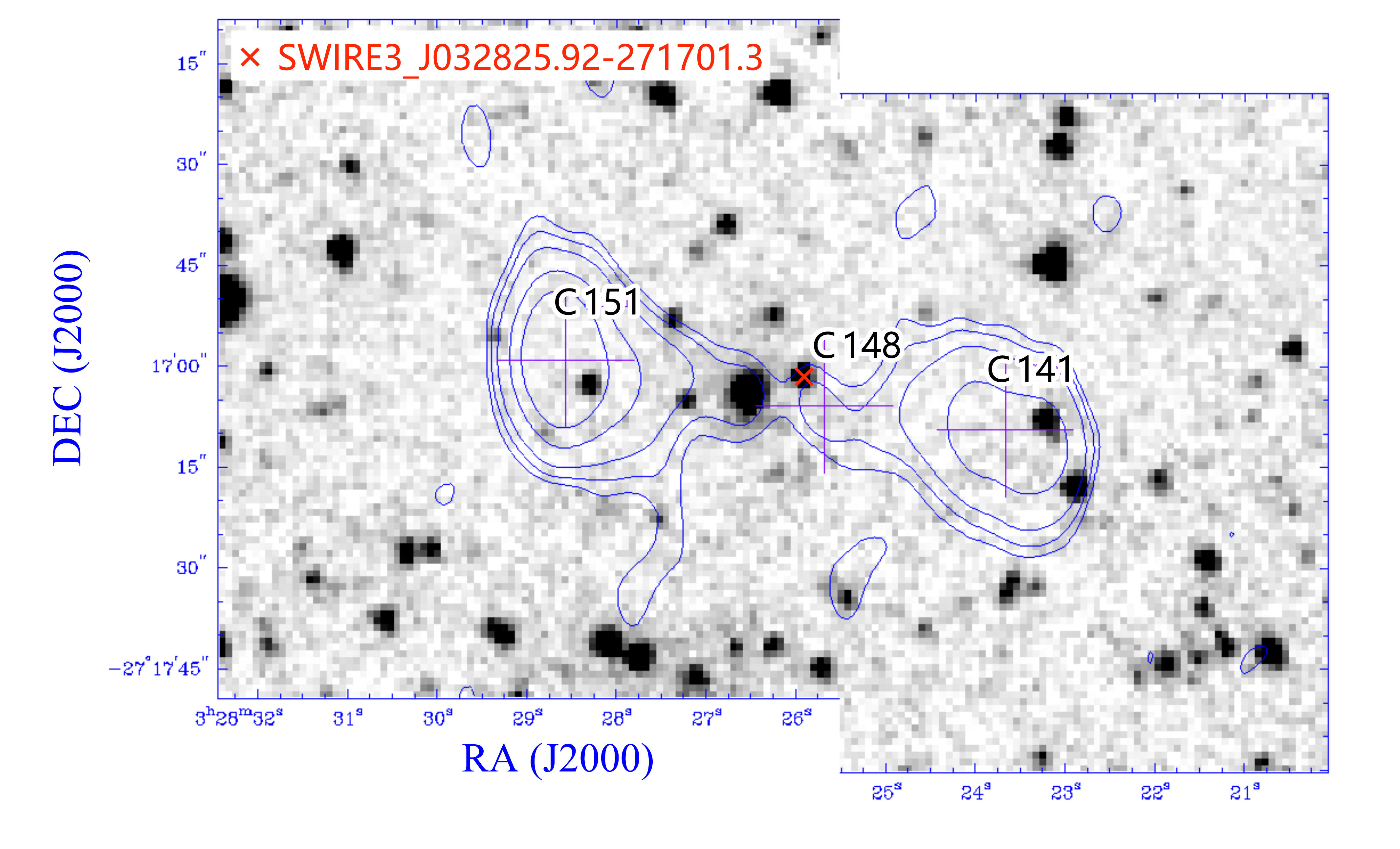}
\end{center}
\caption{The source \texttt{SWIRE3\_J032825.92-271701.3},  with \C{148}--\C{151}--\C{141}, classed as ``complex'' in the reference catalogue and as a triple by ILP. The SWIRE object is marked by a red ``$\times$'' at the center of this figure. In this and subsequent figures, greyscale is the 3.6$\mu$m SWIRE image, and the contours are the radio image from ATLAS. \label{img:C141-C148-C151}
}
\end{figure}

\begin{figure}
\begin{center}
\includegraphics[scale=0.25]{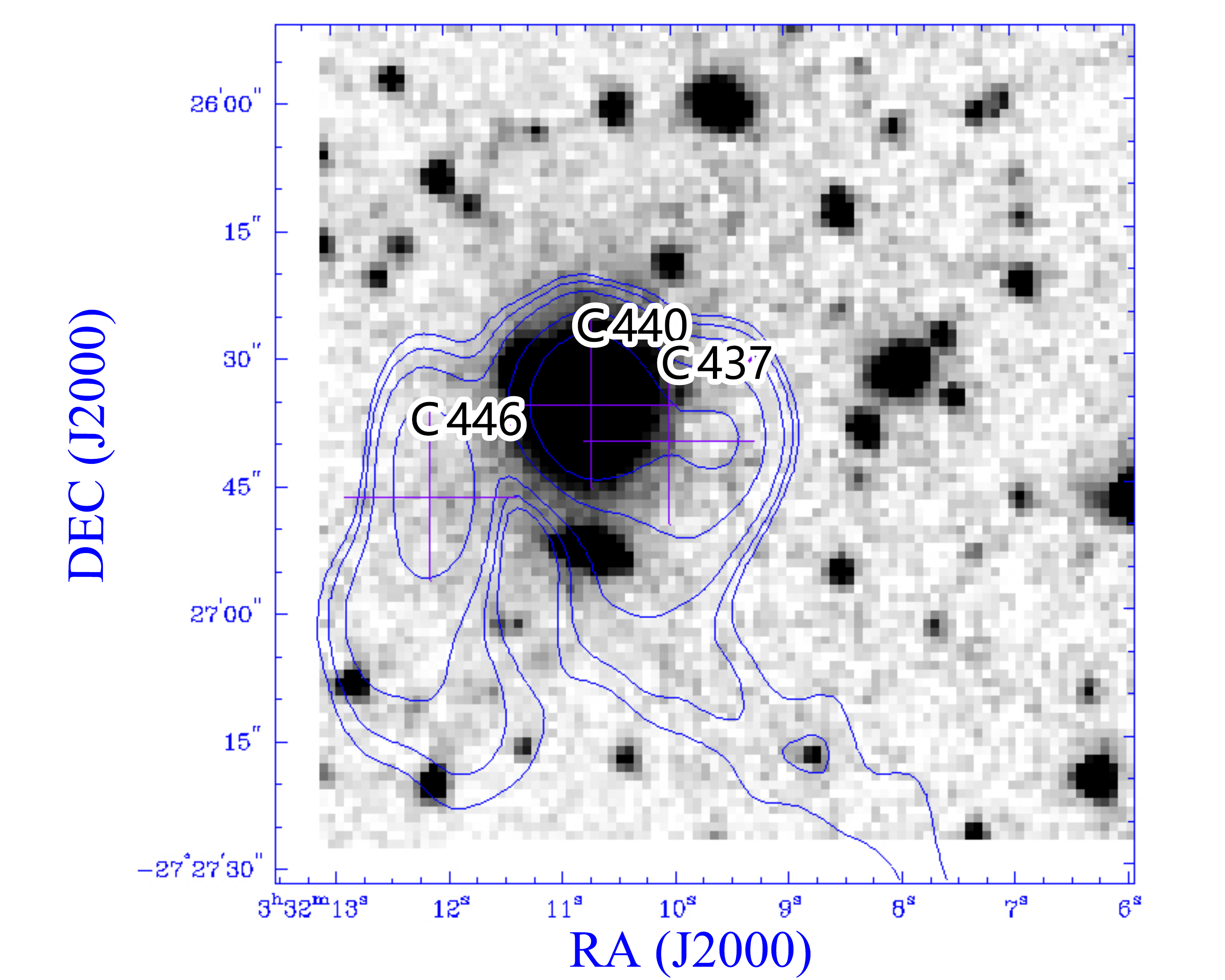}
\end{center}
\caption{
The source \texttt{SWIRE3\_J033210.74-272635.5} with \C{437}--\C{440}--\C{446}, classed as ``complex'' in the reference catalogue and as a triple by ILP.}
\label{img:C437-C440-C446}
\end{figure}

\subsection{Radio Triples}

Since the code finds all radio triples in the reference catalogue, we study the  two extra associations that our approach also identifies:
\begin{enumerate}

\item
The radio components \C{148}--\C{151}--\C{141} (shown in  Figure~\ref{img:C141-C148-C151}) associated with \texttt{SWIRE3\_J032825.92-271701.3} also appear in the reference catalogue but not as a triple,
because there is no central radio component associated with an IR source.
Instead they form complex \#1 with the comment: \textit{``Radio double with connecting jet''}.
The ILP has identified the IR source SWIRE3\_J032825.92-271701.3 as the host galaxy. While this is well separated from the radio component C148, there is low-level radio emission associated with this SWIRE source, so it possible that this is the correct host galaxy, even though it does not seem to be associated with the listed central radio component.

\item
The radio components \C{440}--\C{446}--\C{437} associated with \textsf{SWIRE3\_J033210.74-272635.5} shown Figure~\ref{img:C437-C440-C446} also appear in the reference catalogue as complex \#3 with the comment \textit{``weird complex source with tails. z(g)''}.
It was not classified as a triple in the reference catalogue because to do so would ignore the bent tails, probably resulting from an interaction with an intra-cluster medium. However, here we focus only on the positions of the components, and in those terms this source is correctly classified as a triple.

\end{enumerate}

So, we conclude that, provided only the positions of the components are considered, one of the two extra sources is correctly classified as a triple. So our procedure associates radio triples with a 92\% success rate, with the only failure being a source that closely resembles a triple, although lacking a central core.

\subsection{Radio Doubles}\label{sec:radio_doubles}
Our procedure correctly classified 21 of the 31 doubles, missed 4, and incorrectly identified a further 10 sources as doubles. However, we caution that, as discussed in Section \ref{results} above,  the reference  catalogue can occasionally be unreliable. In some cases below, it is hard to tell, with the given data, whether the Reference catalogue or the ILP is more likely to be correct.

\begin{figure*}
\begin{center}
\includegraphics[clip,trim=15mm 0 15mm 0,scale=0.18]{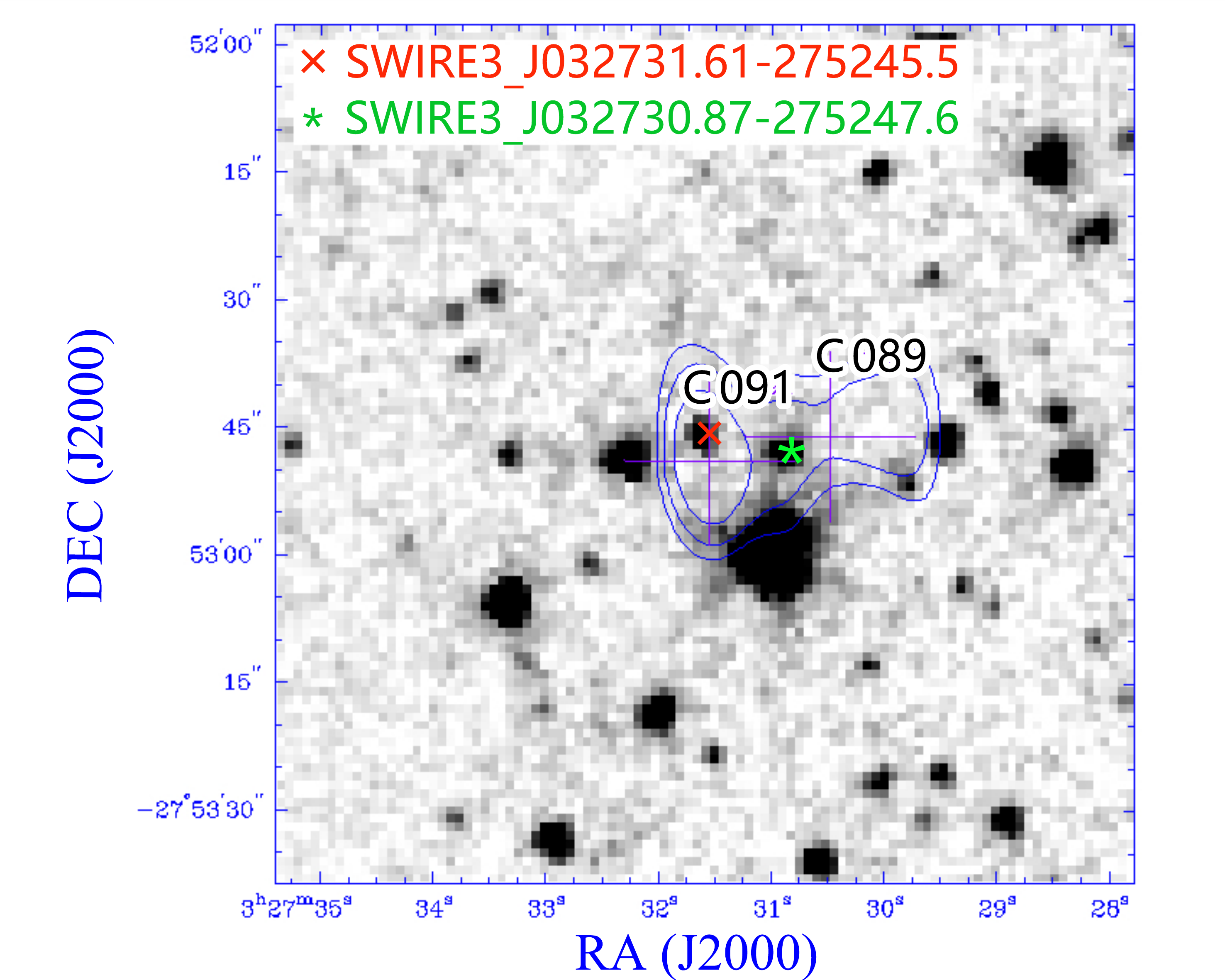}
\includegraphics[clip,trim=15mm 0 15mm 0,scale=0.18]{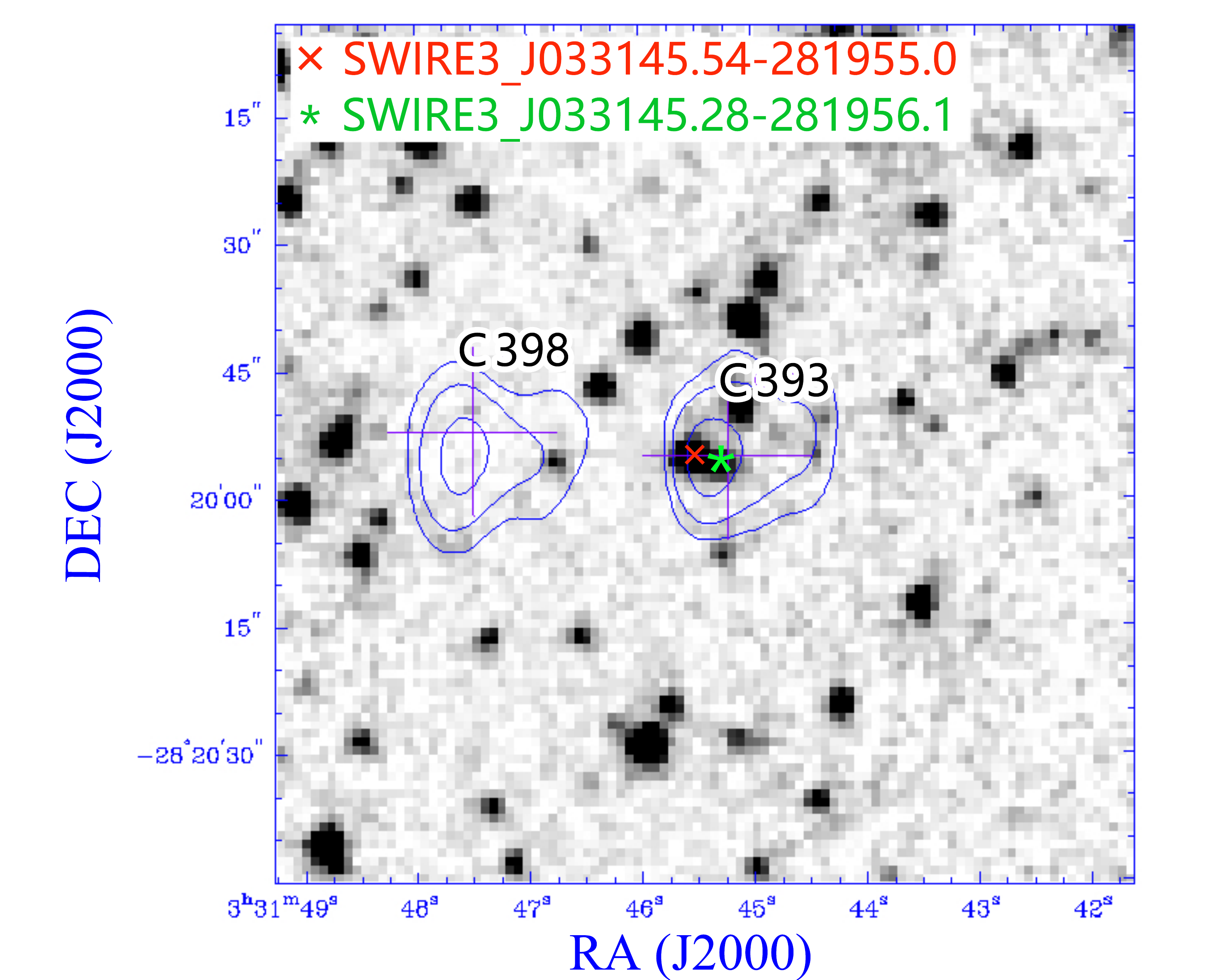}
\includegraphics[clip,trim=15mm 0 15mm 0,scale=0.18]{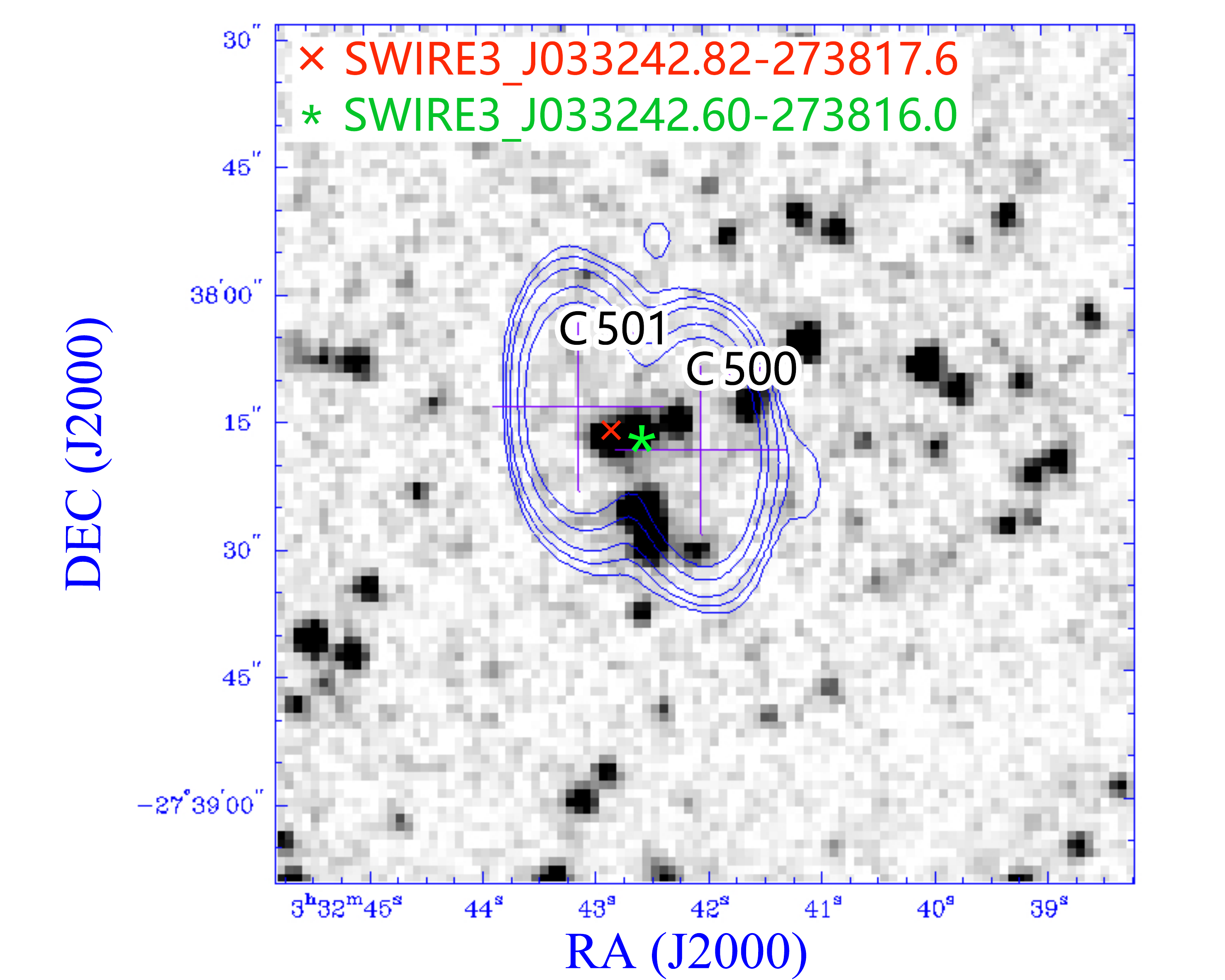}
\includegraphics[scale=0.25]{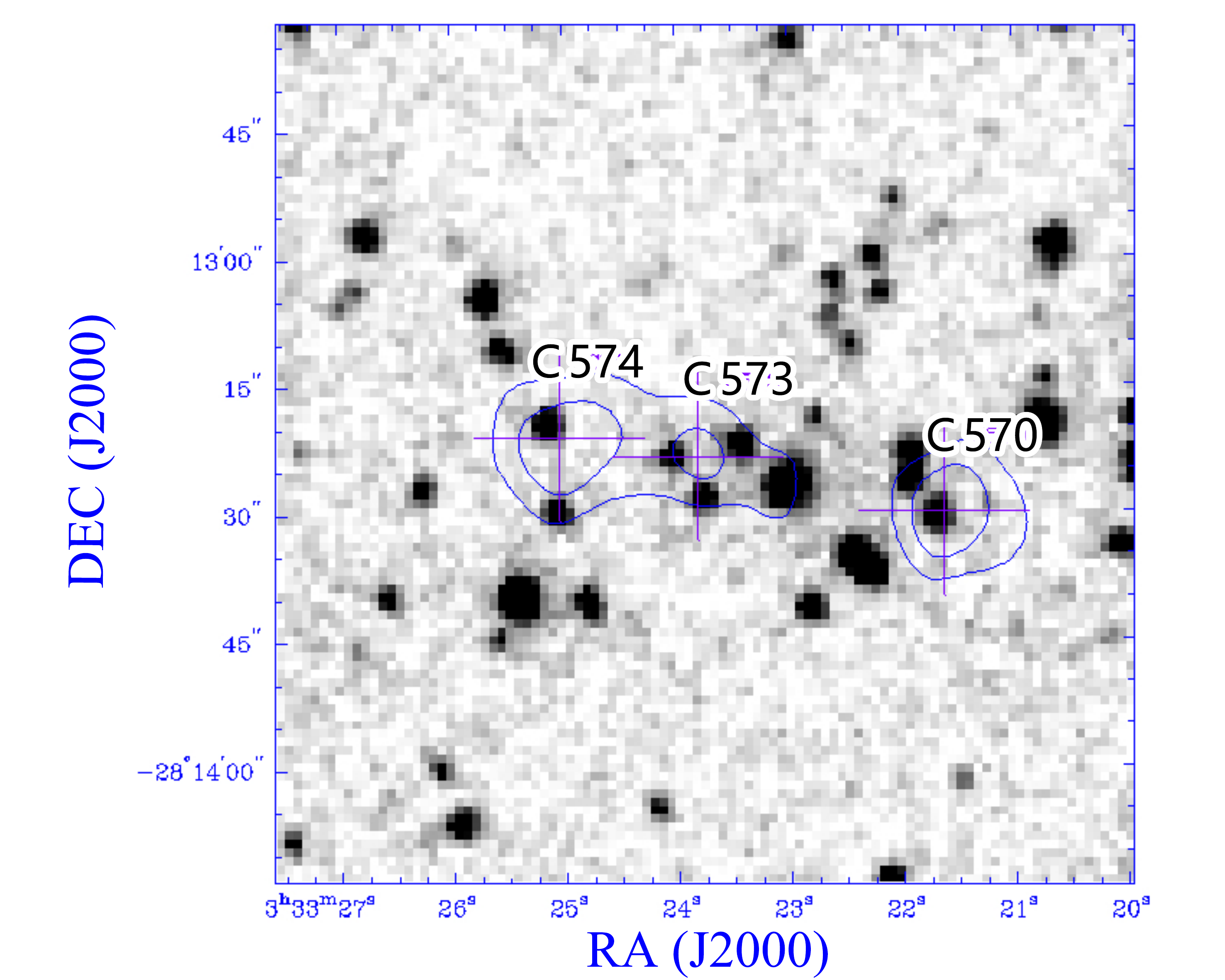}
\includegraphics[scale=0.25]{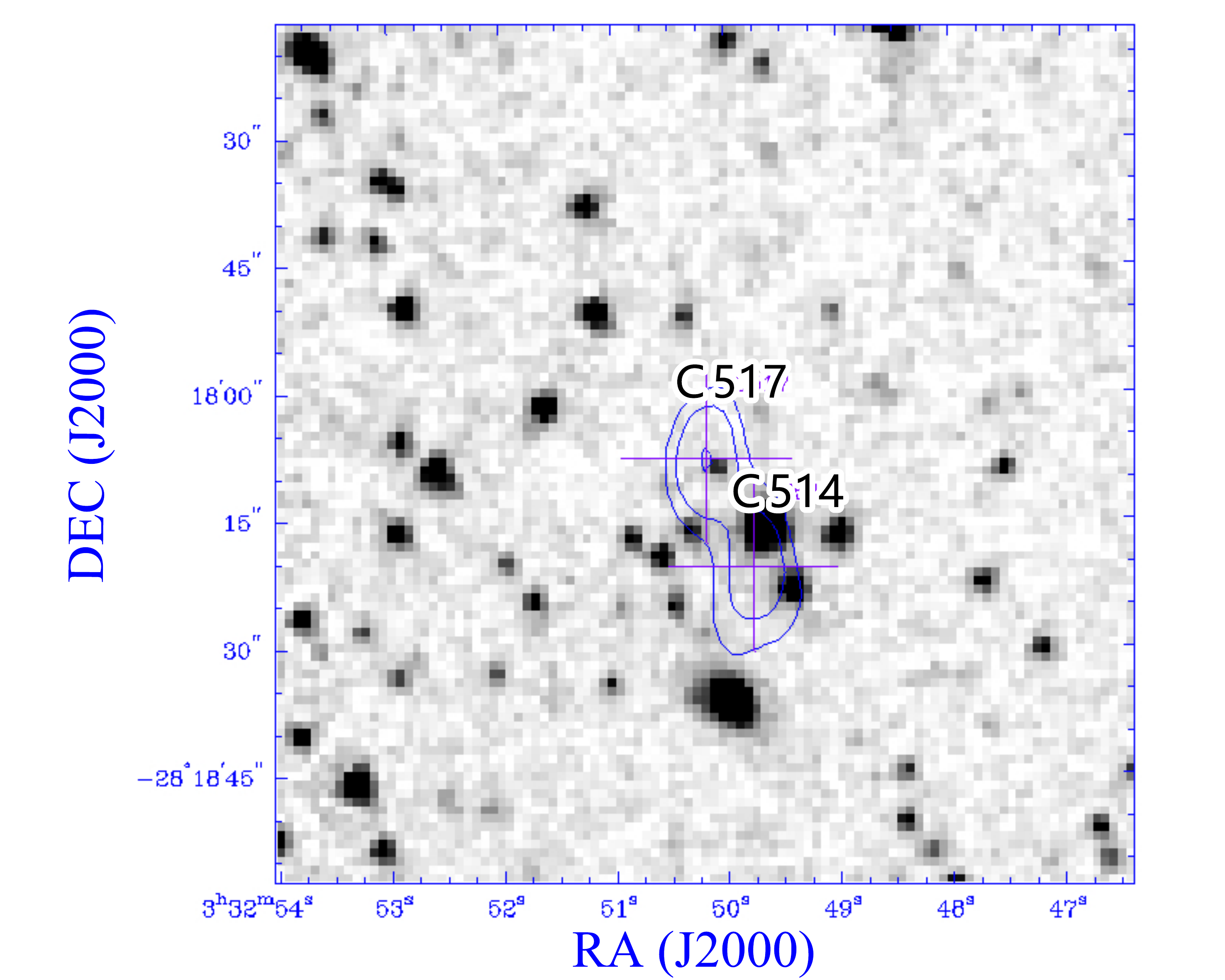}
\end{center}
\caption{
    Radio Doubles discussed in Section \ref{sec:radio_doubles}.
    The first row, from left to right are \C{089}--\C{091}, \C{393}--\C{398} and \C{500}--\C{501}, ILP detects them as doublet as the reference catalogue, but ILP associates them with different IR object.
    Hereafter, red ``$\times$'' represents the infrared choice of reference catalogue, while green ``$\ast$'' indicates the choice of ILP.
    On the bottom left image, ILP classifies \C{573}--\C{574} as two singles, while in the reference catalogue they are a doublet. Conversely, \C{517}--\C{514} on the bottom right image are classified by ILP as a Core-Lobe configuration, but the reference catalogue lists them as two singles.
    } \label{img:radio-doubles}
\end{figure*}

Our procedure misses four of the radio doubles in the reference catalogue: \C{089}--\C{091}, \C{393}--\C{398}, \C{500}--\C{501}, and \C{573}--\C{574}, as shown in Figure~\ref{img:radio-doubles} .

\C{089}--\C{091} with \textsf{SWIRE3\_J032731.61-275245.5} is classified in the reference catalogue as ``complex region, but probably core-jet with C091 core, C089 jet''; ILP classified \C{089}--\C{091} with the other \textsf{SWIRE3\_J032730.87-275247.6} as a ``LOBE--LOBE''.

\C{393}--\C{398} with \textsf{SWIRE3\_J033145.54-281955.0} is classified in the reference catalogue as ``Probably asymmetric radio double''; ILP classified \C{393}--\C{398} with the other \textsf{SWIRE3\_J033145.28-281956.1} as a ``CORE--LOBE''.

\C{500}--\C{501} with \textsf{SWIRE3\_J033242.82-273817.6} is classified in the reference catalogue as ``Radio double''; ILP classified \C{500}--\C{501} with the other \textsf{SWIRE3\_J033242.60-273816.0} as ``LOBE--LOBE'' (i.e. a double), so the only difference here is the choice of the host galaxy.

Interestingly, the first three of them actually appear as extra doubles, but in associations with different SWIRE measurements. Upon visual inspection, these mismatches appear to be a result of limited information in the positions only considered in this paper. It is possible that including brightness or shape measurements might help find the correct associations, but this needs to be studied further.

The fourth double of \C{573}--\C{574} also appears in our catalogue, but it is separated into two singles due to two SWIRE detections close to the radio components, see bottom left image of Figure~\ref{img:radio-doubles}.
The reference  catalogue comment reads \textit{``M-test classifies it as double with C573, but C574 has good ID so C573 is probably a jet''}.

Our code produces seven extra radio doubles: six of them
were classified as ``complex'' as shown in Table~\ref{tbl:5-extra-doublet}.
For example, \#6 was classified as a double in the comments to the reference catalogue, and has been correctly identified as such by the ILP, but it also contains a third component, C774, as an extension to one of the lobes, and is therefore classified as complex in the reference catalogue rather than as a simple double. Thus the ILP's classification as ``double'' in this case is possibly a more useful classification than ``complex''.
\label{C774}

The remaining \C{517}--\C{514} is classified by ILP as a Core-Lobe configuration, but the
reference catalogue
lists them as two singles, as shown the bottom right image of Figure~\ref{img:radio-doubles}.
Given the astrometric uncertainty, it is difficult to tell whether C517 and C514  are associated with the IR sources, so in this case it is hard to tell whether the disagreement represents an error in ILP or an error in the reference catalogue.

\begin{table}
\begin{center}
\caption{Extra doubles found by ILP appear in complexes\label{tbl:5-extra-doublet}}
\begin{tabular}{lll}
\hline\hline
Cplx & Radios Components & Associations by ILP\\
\hline
\multirow{2}{*}{\#2}&\multirow{2}{*}{\C{366}, \C{369}, \C{376}}
& \C{376}--\C{366} as \texttt{Core--Lobe}\\
&& \C{369}--\C{381} as \texttt{Lobe--Lobe}\\
\cline{3-3}
\multirow{2}{*}{\#4}&\multirow{2}{*}{\C{679}, \C{681}, \C{683}}
& \C{679} not in ILP result\\
&& \C{683}--\C{681} as \texttt{Lobe--Lobe}\\
\cline{3-3}
\multirow{2}{*}{\#5}&\multirow{2}{*}{\C{759}, \C{760},\  \C{761}, \C{764}}
& \C{759}--\C{760} as \texttt{Lobe--Lobe}\\
&& \C{761}--\C{764} as \texttt{Core--Lobe}\\
\cline{3-3}
\multirow{2}{*}{\#6}&\multirow{2}{*}{\C{774}, \C{775}, \C{777}}
& \C{774} as extra \texttt{Single}\\
&& \C{775}--\C{777} as \texttt{Lobe--Lobe}\\
\hline\hline
\end{tabular}
\end{center}
\end{table}

\subsection{Radio Singles} \label{sec:radio_singles}

Singles are individual radio components associated with a SWIRE detection.
Out of the 685 single associations in the reference catalogue, our approach identifies 680, with 4 extra associations.

\begin{figure*}
\begin{center}
\includegraphics[scale=0.25]{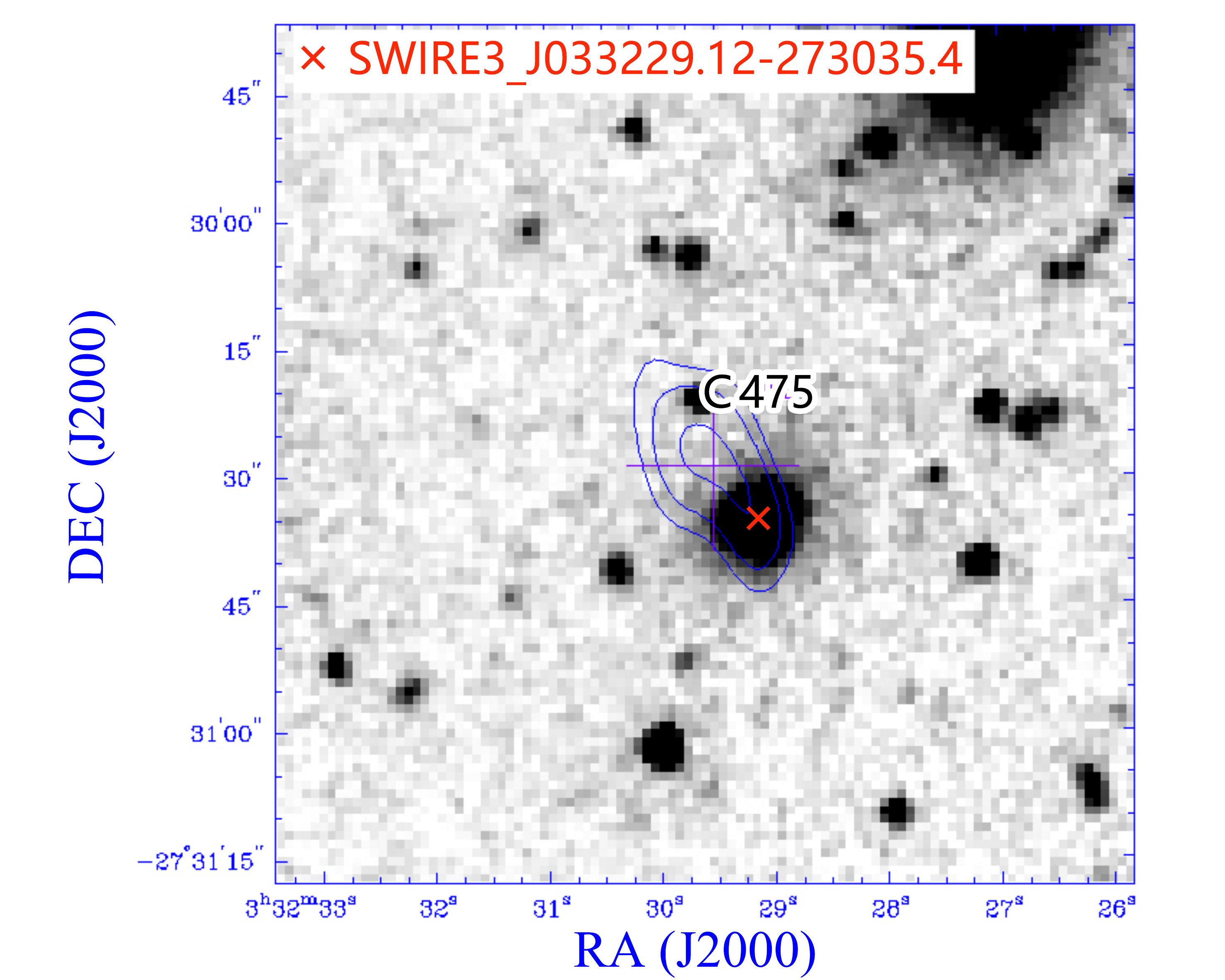}
\includegraphics[scale=0.25]{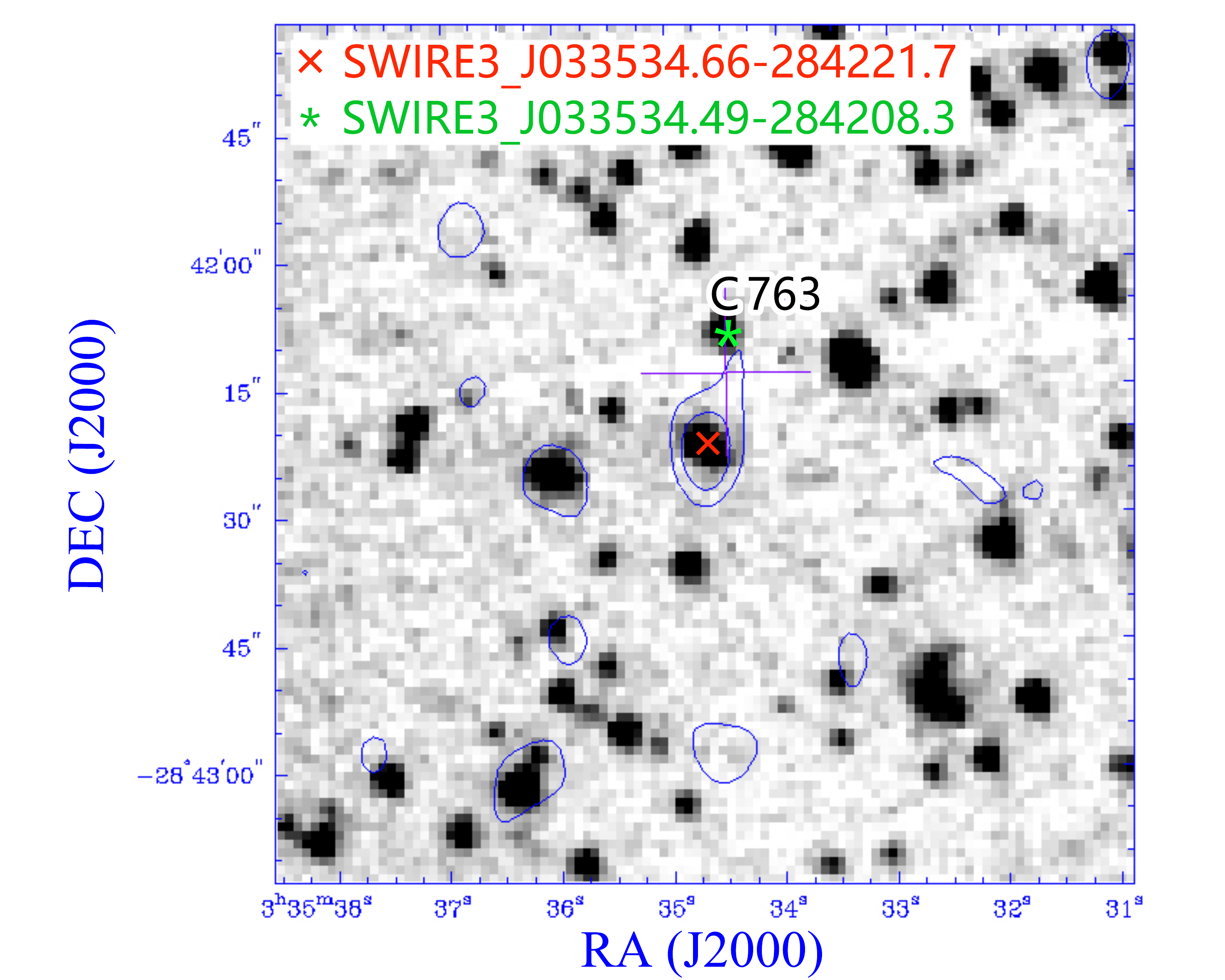}

\end{center}
\caption{
    Radio Single sources discussed in Section \ref{sec:radio_singles}. ILP cannot find an association for \C{475} while it is classified as a jet in the reference catalogue. On the right image, \C{763} is associated with a nearer but weaker \texttt{SWIRE3\_J033534.49-284208.3} in ILP, while in the reference catalogue the IR object is \texttt{SWIRE3\_J033534.66-284221.7}.
    }
    \label{img:radio_single}
\end{figure*}

Out of the five missed (\C{381}, \C{475}, \C{514}, \C{517}, \C{763}), three \C{381}, \C{514}, \C{517} are listed in the reference catalogue   as parts of doubles.
The reference catalogue  comment for \C{475} (see Figure~\ref{img:radio_single})  is \textit{``strong spitzer ID at one end of core-jet src.''}, but this source is too far from that SWIRE source to pass our detection threshold (see Section \ref{practical}), resulting in a low Bayes factor for this association.

Finally, \C{763} (see Figure~\ref{img:radio_single}) is associated with a different SWIRE object from the reference  catalogue, so it is listed as an extra. The ILP associates the radio source with a weak SWIRE detection, instead of a brighter source farther away.  It is interesting that the reference  catalogue identification is associated with the peak of the contours, but the ILP association is with the catalogue position, which is significantly different from the peak of the contours, suggesting an error in the reference catalogue.

There are 4 extra singles, most of them are mentioned above, except for \C{774}, which is part of the complex \#6, which is discussed above in Section \ref{C774}.

\subsection{Multiple solutions}
As discussed in Section \ref{partitions}. the ILP technique generates many slightly different hypotheses, just one of which is chosen as the optimal solution.

\begin{figure}
\begin{center}
\includegraphics[scale=0.25]{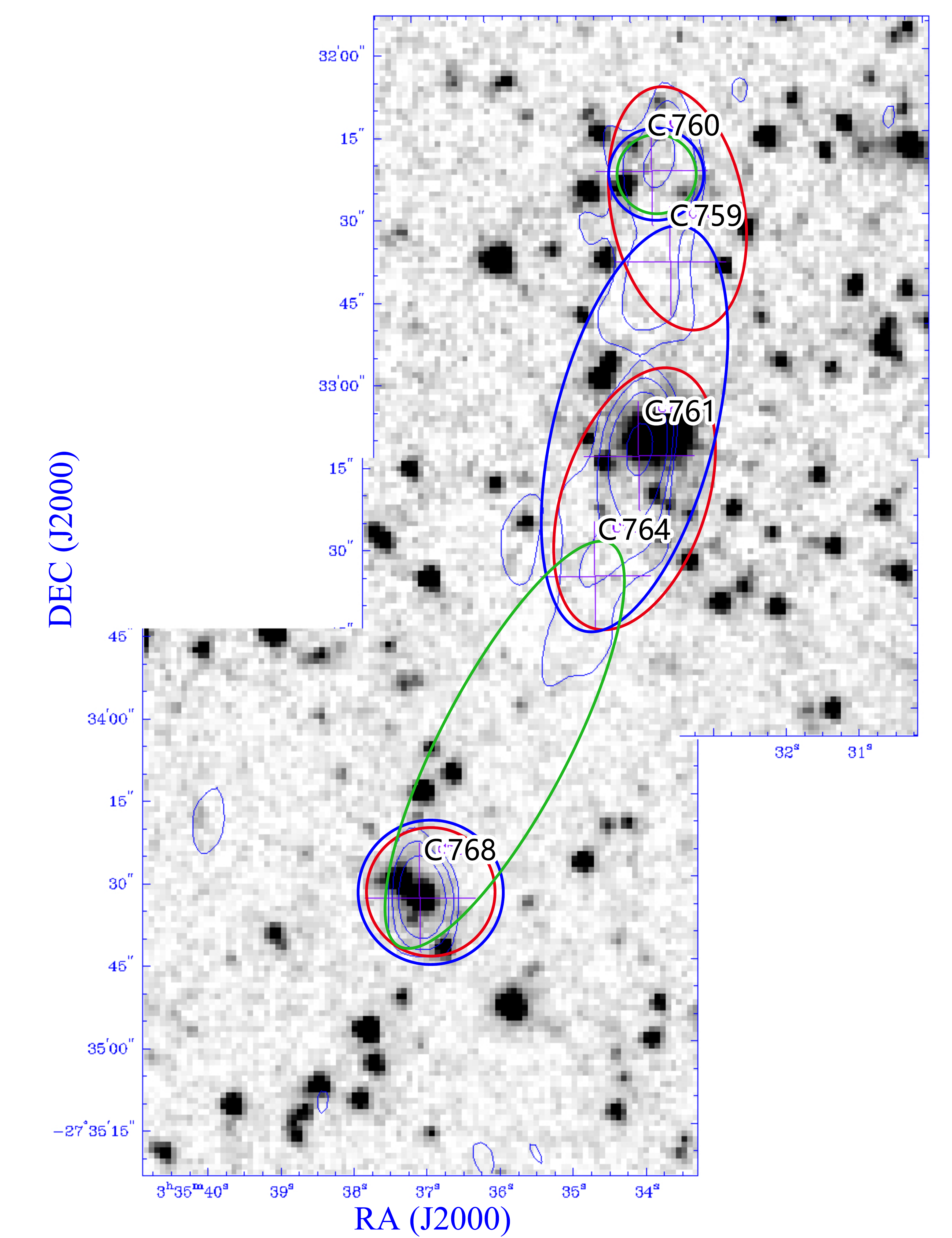}
\end{center}
\caption{An llustration of the difference among ILP solutions. Red assignment is the first solution, Blue is the second, Green is the third. Image details as in Figure \ref{img:C141-C148-C151}
\label{img:compare-ilp-solutions}
}
\end{figure}

In addition to the best solution, our approach can also create similarly good but slightly sub-optimal associations. We studied the first 3 solutions, which show surprisingly little variation.
The difference is due to the assignments of \C{759}, \C{760}, \C{761}, \C{764} and \C{768}, which we illustrate in Figure~\ref{img:compare-ilp-solutions}.
These actually form complex \#5 with comment \textit{``Centre of a linear complex (jets or grav arc?)''}.

The differences are as follows:
In the best solution, \C{759}--\C{760} appears as a Lobe-Lobe combination and \C{761}--\C{764} is a Core-Lobe with \C{768} being single.
In the second solution, lists \C{761}--\C{759}--\C{764} as a radio triple with singles \C{760} and \C{768}.
In the third solution, there \C{761} and \C{759} do not appear, and \C{764}--\C{768} is a Lobe-Lobe, with single \C{760},
as summarized in Table.~\ref{tbl:ilp-sol-difference}

\begin{table}
\begin{center}
\caption{ILP solutions involving detection in \mbox{complex~\#5}}
\label{tbl:ilp-sol-difference}
\begin{tabular}{cc|c|c|c|c}
\hline\hline
Solution & \C{760} & \C{759} & \C{761} & \C{764} & \C{768} \\
\hline
1 & \multicolumn{2}{c|}{\color{red}Lobe - Lobe}& \multicolumn{2}{c|}{\color{blue}Core - Lobe} & {\color{black}Single} \\
\cline{2-6}
2 & {\color{red}Single} & \multicolumn{3}{c|}{\color{blue}Lobe - Core - Lobe} & {\color{black}Single} \\
\cline{2-6}
3 & {\color{red}Single} & \multicolumn{2}{c|}{} &\multicolumn{2}{c}{\color{blue}Lobe - Lobe} \\
\hline\hline
\end{tabular}
\end{center}
\end{table}

Our algorithm will, of course, perform correctly only if there are no errors in the catalogue. In particular, we assume that the catalogue lists one position per radio component, which is true for the data being considered here, and generally true for the source finder (Selavy) used for the EMU surveys \citep{norris11}. However, we are aware that some source finders can generate several Gaussian components for one radio component. While we have not tried the ILP algorithm on catalogues produced by such source finders, we suspect a  pre-processing stage may be need to consolidate the listed components into one per radio component.

\section{Conclusions}\label{conlusions}

Our new method for radio cross-identification has two major improvements over previous automatic attempts. Its flexibility to model the morphology of bent radio lobes is proven vital to recover all triples in \citet{norris06}.
The biggest conceptual changes is in the creating of the associations, which are not chosen to be globally optimal or the entire matched catalogue. The success of this combinatorial optimization approach based on the marginal likelihoods of hypothesised objects is due to our efficient formulation of the problem as integer linear programming, following ideas put forward by \citet{shi2019}.
The results from this automated approach match closely the current state of the art catalogue manually created by experts \citep{norris06}.

For the morphological modeling, each parameter has priors motivated by observations and their uncertainties. The range of values they take appear to be physically meaningful and in line with our current understanding. That said, we expect great improvements in the future when more data will be available. In such setting, a hierarchical approach will be possible to simultaneously learn the population distributions of the parameters and finding the best associations.
Additional improvements are expected from the inclusion of brightness and shape measurements but their role and effect are to be studied further.

Also, prior information on possible matched catalogues will play an important role. The current method is essentially based on marginal likelihoods, and does not take into account the frequency of various radio configurations in the universe. A first we take in this position is the inclusion of a prior that prefers smaller number of objects in the final catalogue, which helps with creating associations in situations where other hypotheses are equally likely. In practice, such a bias is preferred because based on additional data (e.g., during visual inspections) the associated detections could potentially be broken apart easily, but going the other way is impossible.

While the optimal algorithm processes the entire data collection in one large ILP solver, we note that in practice the candidate associations will most likely not build a single connected component. Using standard graph analytics, we can split the problem up by finding the actual connected components and process them separately. This is expected to make a huge different in the wall-clock time due to the combinatorial nature of the problem. In our study, we found that such pre-processing was not needed because the entire analysis took only a few minutes.

We are looking forward to running the ILP algorithm on new datasets from  large new surveys such as those on ASKAP \citep{norris11} and MWA \citep{white20}, and comparing the results with other algorithms, particularly ML algorithms. The ASKAP data are similar in sensitivity and resolution to the ATLAS data used here, and so we expect to be able to use ILP on ASKAP data with only modest retraining. MWA data, which is quite different from the sample used in this paper, appear broadly similar in the  statistical characteristics of the radio data, but will pose a great challenge because of the large difference in the resolution of the radio and IR data, so significant retraining will be necessary. It is difficult to predict how ILP will fare on these datasets, and so it is important to experiment.

In preparation for these upcoming surveys, we are currently working with much larger catalogues containing 162 thousand radio detections and 329 million IR sources for which a depth-first cluster finding yields over a factor a 100 speed-up, which takes 66 minutes to run the whole cross-matching.

\section*{Acknowledgements}
DF acknowledges support from Joint Research Fund in Astronomy (U1931132, U1731243) under cooperative agreement between the National Natural Science Foundation of China (NSFC) and Chinese Academy of Sciences (CAS), the CAS Scholarship, and the China National Astronomical Data Center (NADC).
TB gratefully acknowledges support from National Science Foundation (NSF) grants AST-1412566, AST-1814778 and AST-1909709, as well as NASA via STSCI-51038.
AB gratefully acknowledges support from National Science Foundation (NSF) grant CMMI1452820 and Office of Naval Research (ONR) grant N000141812096.


\section*{Data availability}
The data underlying this article are available in China-VO PaperData Repository, at \url{https://doi.org/10.12149/101026}.






\bsp	
\label{lastpage}
\end{document}